\documentclass[aps,prl,twocolumn,superscriptaddress]{revtex4-2}
\usepackage{graphicx}
\usepackage{amssymb}
\usepackage{amsmath}
\usepackage{booktabs}
\usepackage{multirow}
\usepackage{bm}
\usepackage{dsfont}
\usepackage{xcolor}
\usepackage{soul}
\usepackage{CJKutf8}
\usepackage[colorlinks=true,linkcolor=blue,citecolor=blue,urlcolor=blue]{hyperref} 
\newcommand{\fm}{{\rm fm}}

\newcommand{\MeV}{{\rm MeV}}
\newcommand{\GeV}{{\rm GeV}}
\usepackage{cleveref} %use \cref{}
    \crefformat{figure}{Fig.~#2#1#3}
    \crefformat{table}{Table~#2#1#3}
    \crefformat{equation}{Eq.~(#2#1#3)}
    \crefformat{section}{Sect.~#2#1#3}
    \crefformat{appendix}{App.~#2#1#3}
\allowdisplaybreaks
\begin{document}

%%%%%%%%%%%%%%%%%%%%%%%%%%%%%%%%%%%%%%%%%%%%%%
%%%%%%%%%%%%%%%%%%%%%%%%%%%%%%%%%%%%%%%%%%%%%%
\title{$\omega$ Meson from lattice QCD}
%%%%%%%%%%%%%%%%%%%%%%%%%%%%%%%%%%%%%%%%%%%%%%
%%%%%%%%%%%%%%%%%%%%%%%%%%%%%%%%%%%%%%%%%%%%%%

%%%%%%%%%%%%%%%%%%%%%%%%%%%%%%%%%%%%%%%%%%%%%%
\author{Haobo~Yan (\begin{CJK*}{UTF8}{gbsn}燕浩波\end{CJK*})}
\email{haobo@stu.pku.edu.cn}
\affiliation{School of Physics, Peking University, Beijing 100871, China}
\affiliation{Helmholtz-Institut f\"ur Strahlen- und Kernphysik (Theorie) and Bethe Center for Theoretical Physics,  Universit\"at Bonn, 53115 Bonn, Germany}
%%%%%%%%%%%%%%%%%%%%%%%%%%%%%%%%%%%%%%%%%%%%%%
\author{Maxim~Mai}
\email{maxim.mai@faculty.unibe.ch}
\affiliation{Helmholtz-Institut f\"ur Strahlen- und Kernphysik (Theorie) and Bethe Center for Theoretical Physics,  Universit\"at Bonn, 53115 Bonn, Germany}
\affiliation{Albert Einstein Center for Fundamental Physics, Institute for Theoretical Physics, University of Bern, Sidlerstrasse 5, 3012 Bern, Switzerland}
\affiliation{Institute for Nuclear Studies and Department of Physics, The George Washington University, Washington, DC 20052, USA}
%%%%%%%%%%%%%%%%%%%%%%%%%%%%%%%%%%%%%%%%%%%%%%
\author{Marco~Garofalo}
\email{garofalo@hiskp.uni-bonn.de}
\affiliation{Helmholtz-Institut f\"ur Strahlen- und Kernphysik (Theorie) and Bethe Center for Theoretical Physics,  Universit\"at Bonn, 53115 Bonn, Germany}
%%%%%%%%%%%%%%%%%%%%%%%%%%%%%%%%%%%%%%%%%%%%%%
\author{Ulf-G.~Mei{\ss}ner}
\email{meissner@hiskp.uni-bonn.de}
\affiliation{Helmholtz-Institut f\"ur Strahlen- und Kernphysik (Theorie) and Bethe Center for Theoretical Physics,  Universit\"at Bonn, 53115 Bonn, Germany}
\affiliation{Institute for Advanced Simulation (IAS-4), Forschungszentrum J\"ulich, 52425 J\"ulich, Germany}
\affiliation{Peng Huanwu Collaborative Center for Research and Education, Beihang University, Beijing 100191, China}
%%%%%%%%%%%%%%%%%%%%%%%%%%%%%%%%%%%%%%%%%%%%%%
\author{Chuan~Liu}
\email{liuchuan@pku.edu.cn}
\affiliation{School of Physics, Peking University, Beijing 100871, China}
\affiliation{Center for High Energy Physics, Peking University, Beijing 100871, China}
\affiliation{Collaborative Innovation Center of Quantum Matter, Beijing 100871, China}
%%%%%%%%%%%%%%%%%%%%%%%%%%%%%%%%%%%%%%%%%%%%%%
\author{Liuming~Liu}
\email{liuming@impcas.ac.cn}
\affiliation{Institute of Modern Physics, Chinese Academy of Sciences, Lanzhou 730000, China}
\affiliation{University of Chinese Academy of Sciences, Beijing 100049, China}
%%%%%%%%%%%%%%%%%%%%%%%%%%%%%%%%%%%%%%%%%%%%%%
\author{Carsten~Urbach}
\email{urbach@hiskp.uni-bonn.de}
\affiliation{Helmholtz-Institut f\"ur Strahlen- und Kernphysik (Theorie) and Bethe Center for Theoretical Physics,  Universit\"at Bonn, 53115 Bonn, Germany}
%%%%%%%%%%%%%%%%%%%%%%%%%%%%%%%%%%%%%%%%%%%%%%
% \date{\today}

%%%%%%%%%%%%%%%%%%%%%%%%%%%%%%%%%%%%%%%%%%%%%%
%%%%%%%%%%%%%%%%%%%%%%%%%%%%%%%%%%%%%%%%%%%%%%
\begin{abstract}
Many excited states in the hadron spectrum have large branching ratios to three-hadron final states. Understanding such particles from first principles QCD requires input from lattice QCD with one-, two-, and three-meson interpolators as well as a reliable three-body formalism relating finite-volume spectra at unphysical pion mass values to the scattering amplitudes at the physical point. In this work, we provide the first-ever calculation of the resonance parameters of the $\omega$ meson from lattice QCD, including an update of the formalism through matching to effective field theories. The main result of this pioneering study, the pole position of the $\omega$ meson at $\sqrt{s_\omega}= (778.0(11.2)-i\,3.0(5) )\,\mathrm{MeV}$, agrees reasonably well with experiment. In addition we provide an estimate of the $\omega-\rho$ mass difference as $29(15)\,\mathrm{MeV}$.
\end{abstract}
%%%%%%%%%%%%%%%%%%%%%%%%%%%%%%%%%%%%%%%%%%%%%%
%%%%%%%%%%%%%%%%%%%%%%%%%%%%%%%%%%%%%%%%%%%%%%
\maketitle

%%%%%%%%%%%%%%%%%%%%%%%%%%%%%%%%%%%%
\noindent\emph{Introduction}--%
%%%%%%%%%%%%%%%%%%%%%%%%%%%%%%%%%%%%
Quantum Chromodynamics (QCD), the theory of the strong interactions, not only explains the binding of quarks and gluons to protons and neutrons, which represent most of the visible matter around us, but also the full spectrum of the so-called hadrons. 
It consists in general of baryon and meson states, most of which are actually resonances.
The $\omega(782)$ meson plays a special role in this hadron spectrum. First, it is the lightest hadron that features a strong, isospin conserving decay into three particles in the final state, $\omega \to 3\pi$. Second, within the vector dominance picture of the photon-nucleon interactions, it dominates the isoscalar response~\cite{Sakurai:1960ju, Feynman:1973xc} and combined with the topological soliton picture of the nucleon, it allows one to explain the difference in the baryonic charge and the isoscalar electric radius~\cite{Meissner:1986ka,Kaiser:2024vbc}. Third, in the one-boson-exchange picture of the nucleon-nucleon interaction, it generates the observed repulsion at distances below $1\,\fm$; see,  e.g.,~\cite{Erkelenz:1974uj,BJNN}. Fourth, due to strong isospin violation, it mixes with the $\rho(770)$ meson leading to marked effects in the pion vector form factor, see, e.g.,~\cite{Barkov:1985ac,OConnell:1995nse}. 
Finally, the $\omega-\rho$ mass splitting is phenomenologically interesting, for instance for the anomalous magnetic moment of the muon~\cite{FermilabLattice:2021hzx,Hoferichter:2023sli,Stoffer:2023gba}, or recently also in the context of dark matter and so-called mirror matter~\cite{Hippert:2021fch,Hippert:2022snq}. For all these reasons, a first-principles calculation of this intriguing state based on QCD is called for.

The by now standard approach for such a nonperturbative calculation is represented by lattice QCD, where space-time is discretized, and the Euclidean path integral is estimated using Markov Chain Monte Carlo methods. While lattice QCD has already addressed systematically the lowest resonances, the $f_0(500)$~\cite{Alford:2000mm,Prelovsek:2010kg,Fu:2012gf,Fu:2013ffa,Briceno:2016mjc,Doring:2016bdr,Liu:2016cba,Fu:2017apw,Guo:2018zss,Briceno:2017qmb} and the $\rho(770)$~\cite{Guo:2016zos,CP-PACS:2007wro,Bali:2015gji,Lang:2011mn,Feng:2010es,Fischer:2020yvw,Pelissier:2012pi,Erben:2019nmx,CS:2011vqf,Wilson:2015dqa,Bulava:2016mks,Fu:2016itp,Sun:2015enu,Alexandrou:2017mpi,Andersen:2018mau,Akahoshi:2021sxc,ExtendedTwistedMass:2019omo,Boyle:2024hvv,Boyle:2024grr}, which decay into two pions in the final state (for a review see~\cite{Mai:2022eur}), there are only investigations of repulsive three-body systems~\cite{Mai:2018djl, Blanton:2019vdk, Culver:2019vvu, Hansen:2020otl, Fischer:2020jzp,Alexandru:2020xqf,Draper:2023boj}, and only one exploratory lattice investigation of the $a_1(1260)$ axial meson decaying into three pions available so far~\cite{Mai:2021nul}. In particular, there is no calculation of the complex pole position of the $\omega$ meson available, because it decays predominantly to three pions in the final state. The reason is that only in the last decade the required formalism for such types of lattice computations has become available, see recent reviews~\cite{Hansen:2019nir, Mai:2021lwb}. Using one of the three state-of-the-art formalisms~\cite{Mai:2017bge}, we report here on the first lattice calculation of the $\omega(782)$ meson, thus filling in the gap mentioned before by providing the complex energy, namely the mass and the width of the $\omega$. Owing to its three-pion decay, where two pions can form a $\rho$~\cite{Gell-Mann:1962hpq}, the $\omega$ cannot be considered in isolation, and we thus rely on chiral Lagrangians with vector mesons (for a review, see~\cite{Meissner:1987ge}) in the analysis of the $\omega$ self-energy. In particular, we use effective field theory for the extrapolation to the physical pion mass value.

\bigskip
%%%%%%%%%%%%%%%%%%%%%%%%%%%%%%%%%%%%
\noindent\emph{Lattice computation}---%
%%%%%%%%%%%%%%%%%%%%%%%%%%%%%%%%%%%%
The gauge configurations used in this work were generated by the CLQCD Collaboration with $N_{\text{f}} = 2+1$ flavors of dynamical quarks using the tadpole-improved tree-level Symanzik gauge action and tadpole-improved tree-level Wilson clover fermions~\cite{Hu2024}. The results presented here are based on four ensembles at the same lattice spacing $0.07746(18) \,\fm$, with two pion masses $M_\pi \approx 208$ and $305\,\MeV$, and two volumes each. The details of the ensembles are listed in \cref{tab:ens}. Specifically, the two ensembles, F32P21/F48P21 (F32P30/F48P30), share the same pion mass and lattice spacing but differ in volume.
%, which provide more kinematic points and, thus, a more precise determination of the scattering parameters.

The lattice discretization reduces the continuum rotational symmetry to the cubic symmetry group $O_h$ in the rest frame. Therefore, operators satisfying specific transformation laws of the cubic group are constructed to interpolate the $\rho$ and the $\omega$ mesons. This study focuses on the irreducible representation (irrep) $T_1^-$ for both isovector $\pi\pi$ and isoscalar $\pi\pi\pi$ in the rest frame, where the $\pi\pi$ system predominantly involves the $\rho$ in the \textit{P} wave and $\pi\pi\pi$ houses the $\omega$. The constructed operators include types with a single meson, two mesons, and three mesons, projected to the proper isospin and the $T_1^-$ irrep. For an efficient tool for operator construction, see \verb|OpTion|~\cite{Yan2024}. We emphasize that it is necessary to have all three types of operators to overlap with the dynamical channels and the $\omega$, and obtain reliable and precise energy spectra with minimal pollution from the higher energy region. The detailed form of the operators we used can be found in the Supplemental Material~\cite{supp}\nocite{Garcia-Martin:2011nna, Pelaez:2004xp, Colangelo:2001df}. In order to extract the finite-volume spectra, the correlation matrices of a wide range of operators $O_i$, $C_{i j}(t)= \langle O_i^{}t) O_j^{\dagger}(0) \rangle_T$ are diagonalized by solving a generalized eigenvalue problem~\cite{Michael:1982gb,Luscher:1990ck,Blossier:2009kd,Fischer:2020bgv}; see details in~\cite{supp}.
Lattice energy levels $aE$ are extracted from the exponential decay of the principal correlators in Euclidean time. We note in passing that due to exact isospin symmetry in our lattice calculation the channel $\omega\to\pi^+\pi^-$ is forbidden, but accounts only for about 2\% of the $\omega$ decays in total. We also note that there is mixing with the $\varphi$ meson, which is, however, too high in energy to play a role in our analysis.

\begin{figure}[htbp]
\centering
\raisebox{-0.5\height}{\includegraphics[width=0.21\columnwidth]{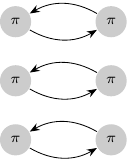}}
\quad
\raisebox{-0.5\height}{\includegraphics[width=0.21\columnwidth]{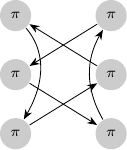}}
\quad
\raisebox{-0.5\height}{\includegraphics[width=0.21\columnwidth]{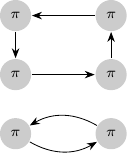}}
\quad
\raisebox{-0.5\height}{\includegraphics[width=0.21\columnwidth]{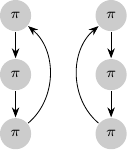}}
\newline\newline\newline
\raisebox{-0.5\height}{\includegraphics[width=0.21\columnwidth]{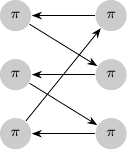}}
\hspace{1cm}
\raisebox{-0.5\height}{\includegraphics[width=0.21\columnwidth]{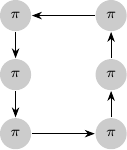}}
\hspace{1cm}
\raisebox{-0.5\height}{\includegraphics[width=0.21\columnwidth]{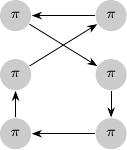}}
\caption{A selection of diagram topologies for $I=0$ for $\pi\pi\pi \to \pi\pi\pi$. All diagrams with the same source and sink are permutations and recombination of these topologies. The other topologies can be found in the Supplemental Material~\cite{supp}.}
\label{fig:topologies}
\end{figure}

The number of quark contraction diagrams emerging in the construction of the relevant correlators grows factorially with the number of scattered particles. For instance, there are nine diagrams for $\pi\pi \to \pi\pi~(I=0)$ when both sink and source are two-body operators, but $202$ in $\pi\pi\pi \to \pi\pi\pi~(I=0)$, with the topologies depicted in~\cref{fig:topologies}
%. The remaining topologies can be found in
and the Supplemental Material~\cite{supp}. Besides the large number of diagrams, most of the diagrams include disconnected quark annihilation subdiagrams, which are difficult to calculate and induce a poor signal. Therefore, we employ the distillation method~\cite{Peardon2009} to compute all-to-all quark perambulators, and construct $C_{ij}$ from these.

%%%%%%%%%
%%%%%%%%%
\begin{table}[t]
\addtolength{\tabcolsep}{6pt}
\begin{tabular}{cccc}
\toprule
Ensemble & Volume & $M_{\pi}/\MeV$ & $N_{\rm confs}$ \\
\midrule
F32P21 & $32^3 \times 64$ & $206.8(2.1)$ & $459$ \\
F48P21 & $48^3 \times 96$ & $207.58(76)$ & $221$ \\
F32P30 & $32^3 \times 96$ & $303.61(71)$ & $777$ \\
F48P30 & $48^3 \times 96$ & $304.95(49)$ & $201$ \\
\bottomrule
\end{tabular}
\caption{CLQCD gauge configurations~\cite{Hu2024} used in this work with a lattice spacing of $a=0.07746(18)\ \mathrm{fm}$. The quoted errors are purely statistical.}
\addtolength{\tabcolsep}{-6pt}
\label{tab:ens}
\end{table}
%%%%%%%%%
%%%%%%%%%

%%%%%%%%%
%%%%%%%%%
%%%%%%%%%
%%%%%%%%%

\begin{figure}[t]
\centering
\includegraphics[height=6.22cm,trim=0 0 1.1cm 0,clip]{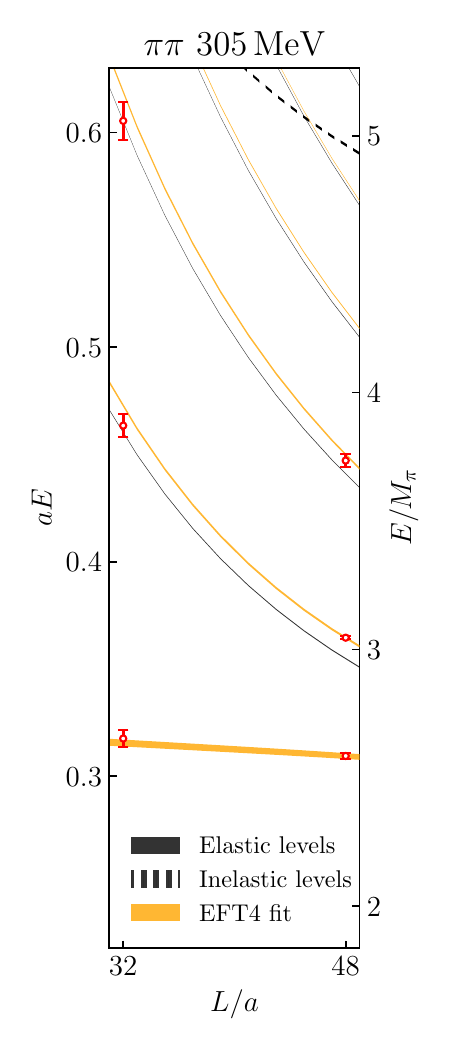}
\includegraphics[height=6.22cm,trim=1.1cm 0 1.1cm 0,clip]{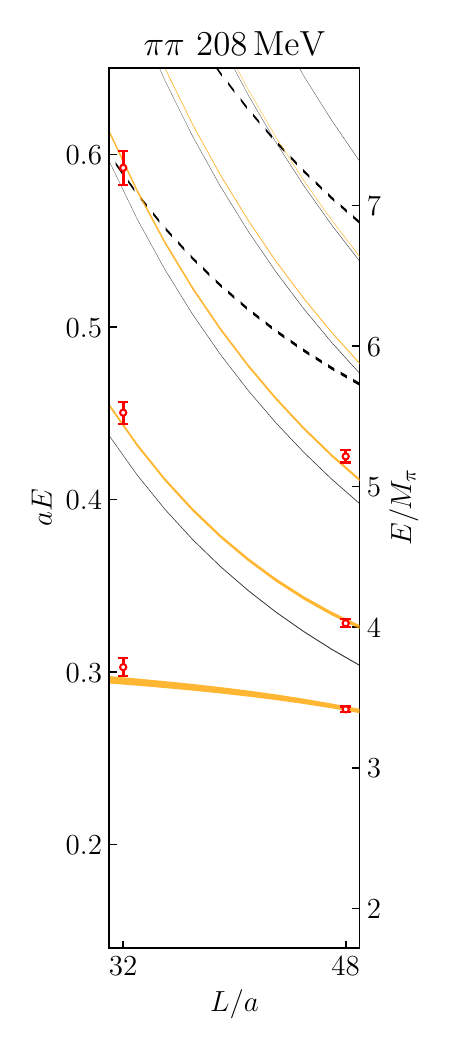}
\includegraphics[height=6.22cm,trim=1.1cm 0 1.1cm 0,clip]{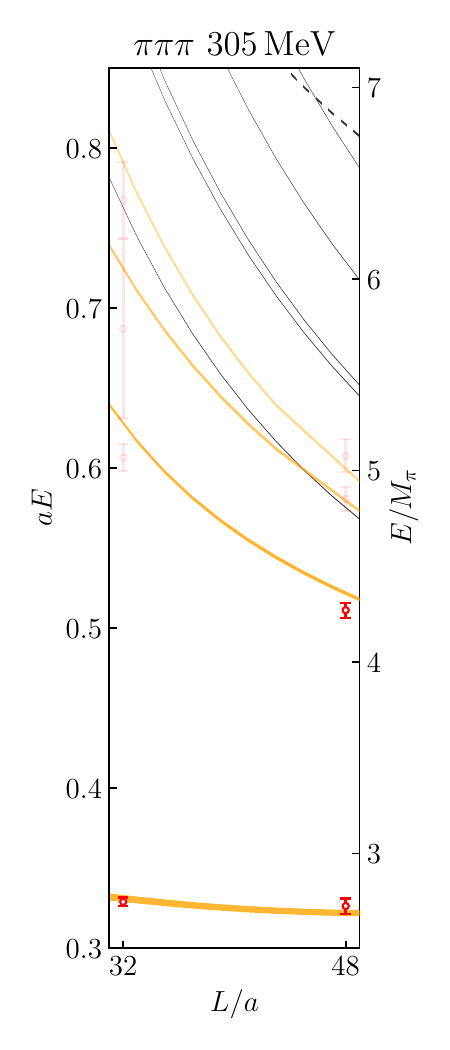}
\includegraphics[height=6.22cm,trim=1.1cm 0 0 0,clip]{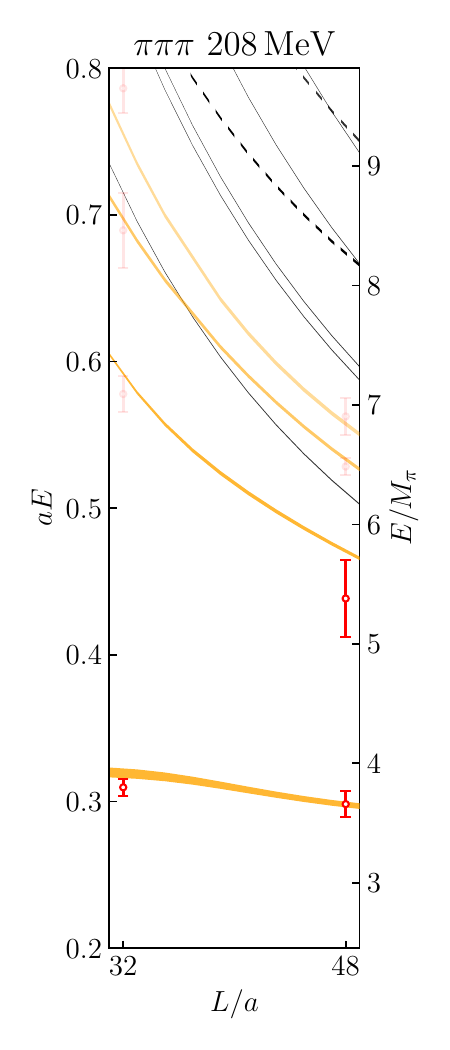}
\caption{
     Finite-volume spectra for $\pi\pi (I=1)$ and $\pi\pi\pi(I=0)$ for heavy and light pion mass. Red points represent the interacting lattice energy levels $aE$ in units of the pion mass, the faded of which are not included in the analysis. Dashed and dot-dashed lines depict the non interacting elastic and inelastic levels. The orange bands are the solutions from the main fit.
    }
\label{fig:spectra}
\end{figure}

The resulting $\pi\pi$ and $\pi\pi\pi$ finite-volume spectra are shown in~\cref{fig:spectra}. The ground levels appear in both $\pi\pi$ and $\pi\pi\pi$ channels below the first non interacting levels indicating strong attraction in both the $\rho$ and $\omega$ channel. In the $\pi\pi\pi$ channel, at $M_{\pi} \approx 208\,\MeV$, the ground states are higher than threshold, indicating a resonance with nonzero phase space to decay; at $M_{\pi} \approx 305\,\MeV$, the ground levels are consistent between the two volumes and lower than the $\pi\pi\pi$ threshold, indicating a bound state.

\bigskip
%%%%%%%%%%%%%%%%%%%%%%%%%%%%%%%%%%%%
\noindent\emph{Quantization conditions and resonance parameter}---%
%%%%%%%%%%%%%%%%%%%%%%%%%%%%%%%%%%%%
The finite-volume spectra discussed above contain two- and three-particle dynamics to be decoded through appropriate quantization conditions. In this work, we utilize the finite-volume unitarity (FVU) approach~\cite{Mai:2017bge} already applied to a variety of three-body systems~\cite{Mai:2018djl, Mai:2019fba, Alexandru:2020xqf, Brett:2021wyd, Mai:2021nul, Garofalo:2022pux, Feng:2024wyg}. It was shown to be equivalent to the other two known three-body formalisms in theory in Ref.~\cite{Jackura:2019bmu} and numerically in Ref.~\cite{Garofalo:2022pux}.

The dominant interaction channel of the $\omega$ system is formalized through the $\pi\rho$ channel in the relative 
\textit{P} wave~\cite{Gell-Mann:1962hpq}. Thus, in the FVU formalism, the three-body finite-volume spectrum is predicted in the center-of-mass frame as a set of three-body energies $E_3=\sqrt{s}$ for which
%%%%%%%%%%
\begin{align}
    \det\{[\tilde K^{-1}(s)-\Sigma^{FV}(s)]E_L-[\tilde B(s)+\tilde C(s)]\}_{\bm{p}'\lambda',\bm{p}\lambda}^\Gamma=0\,,
    \label{eq:QC-tilde}
\end{align}
%%%%%%%%%%
in the plane-wave adn helicity basis (PWH)  $\{\pi(\bm{p})\rho_{\lambda}(-\bm{p}) | \bm{p}L/(2\pi)\in \mathds{Z}^3,\lambda\in\{-1,0,+1\}\}$ while projecting each element of this equation to the $T_1^-$ irrep of the $O_h$ group; see~\cite{Gockeler:2012yj,Mai:2021nul}. The matrices $\tilde B$ (one-pion exchange) and $\Sigma^{FV}$ (self-energy of the $\rho$ system in finite-volume) collect all on shell configurations of the three pions and, therefore, single out all power-law volume dependence of this system. Together with the kinematical factor $E_L$ these matrices are entirely fixed. Contrary to this, the matrices $\tilde K^{-1}$ and $\tilde C$ are volume-independent quantities (up to the neglected $e^{-M_\pi L}$ terms) containing information on the two- and three-body dynamics, respectively. The exponential effects from the small volume were examined by repeating the EFT4 fit excluding the levels from F32P21. We found that the difference from the original fit is of the order of the statistical error, and it is quoted as a systematic error in the Supplemental Material~\cite{supp}. The two-body force provides access to the two-body energy eigenvalues through $\{E_2\in\mathds{R}|\tilde K^{-1}=\Sigma^{FV},\bm{p}^{(\prime)}=\bm{0}\}$ which is equivalent to the usual Lüscher method~\cite{Luscher:1990ux} up to exponentially suppressed terms. While, in general, $\tilde K^{-1}$ and $\tilde C$ are not known,  various choices relevant for the $\rho$ and $\omega$ systems, relying on a generic parametrization and effective field theory are discussed below. 

%%%%%%%%%
%%%%%%%%%
%%%%%%%%%
%%%%%%%%%

Generic method (GEN): the two-body force is parametrized as $[\tilde K^{-1}]_{\bm{p}'\lambda',\bm{p}\lambda}= \delta_{\lambda'\lambda}\delta_{\bm{p}'\bm{p}}\sum_{i=0}^{N}a_i\sigma_{p}^i$ for the two-body invariant mass $\sigma_p:=s+M_\pi^2-2\sqrt{s}\sqrt{\bm{p}^2+M_\pi^2}$ $(E_2=\sqrt{\sigma_{0}})$ and a spectator momentum $\bm{p}$. We found that $N=1$ is entirely sufficient to describe the available lattice input, and this is also mathematically equivalent to the usual Breit-Wigner form. Similarly, the three-body force $\tilde C$ is parametrized through a general expansion in the orbital angular momentum (JLS) basis ($\pi\rho$ in relative \textit{P} wave) $\tilde c_{11}=\frac{c_0}{s-M_\omega^2}+c_1+...$ and then mapped to the PWH~\cite{Chung:1971ri, Sadasivan:2020syi, Feng:2024wyg}. Here again, the order of the expansion depends on the availability and precision of the input. In the current case a two-parameter fit $(c_0,M_\omega)$ turned out as sufficiently flexible.

Effective field theory (EFT): The GEN methodology does not allow 
for chiral extrapolation to the physical point which can be circumvented through EFTs as widely used in the two-body sector~\cite{Mai:2021lwb,Mai:2019pqr} but not yet in studies of three-hadron resonances from the lattice. Still, continuum results on $\omega\to\pi\pi\pi$ and $\rho\to\pi\pi$ have existed for several decades, see the review~\cite{Meissner:1987ge}. Using the results quoted in that review we perform a matching on the level of $2\to2$ and $3\to3$ scattering amplitudes. A somewhat lengthy but straightforward calculation at the tree-level yields
%%%%%%%%%%%%%%%
\begin{align}
    &\left[\tilde K^{-1}\right]_{\bm{p}'\lambda',\bm{p}\lambda}=\delta_{\lambda'\lambda}
    \delta_{\bm{p}'\bm{p}}\frac{\sigma_p-M_\rho^2}{2g^2}\,,\nonumber\\
    &\tilde c_{11}=\frac{6s (M_\rho^2-\sigma_{q}+6g^2f_\pi^2)(M_\rho^2-\sigma_{p}+6g^2f_\pi^2)}{64g^2\pi^3 f_\pi^6(s-M_\omega^2)}\,,
    \label{eq:c11-eft-matching}
\end{align}
%%%%%%%%%%%%%%%
where the latter is expressed in the JLS basis, projected to the PWH basis~\cite{Chung:1971ri, Sadasivan:2020syi, Feng:2024wyg}. Throughout this derivation, we have assumed that two- and three-pion interactions are saturated by the $s$-channel resonance exchange (justified by the narrow width of the $\rho$ and $\omega$ mesons)  and set $g_{\rho\pi\pi}=g_{\omega\rho\pi}=g$ following~\cite{Meissner:1987ge}. Thus far, the matching relations of ~\cref{eq:c11-eft-matching} provide access to the two- and three-body force for given ($g,M_\rho,M_\omega$). The Kawarabayashi-Suzuki-Fayyazuddin-Riazzudin (KSFR) relation~\cite{Riazuddin:1966sw, Kawarabayashi:1966kd} allows one to reduce this set further through $M_\rho=\sqrt{2}gf_\pi$. Indeed, this specifies already a chiral ($M_\pi$) extrapolation through the $f_\pi(M_\pi)$ from chiral perturbation theory~\cite{Gasser:1983yg}. Using the generalized KSFR relation but allowing for a pion-mass independent shift $M_\omega:=M_\rho+\delta=\sqrt{2}gf_\pi+\delta$ defines the EFT2 method referring to free parameters $(g,\delta)$. Abandoning the KSFR relation entirely we define the EFT4 method by $M_\rho=M_V+a\,M_\pi^2,\,M_\omega=M_V+a\,M_\pi^2+\delta$~\cite{Bruns:2004tj,Yu:2023xxf}, leaving us with four free parameters $(g,M_V,a,\delta)$. Clearly, the proposed EFT methods only represent a larger class of EFTs with heavy degrees of freedom~\cite{Meissner:1986ka}. Ultimately, the defined method will be tested against lattice QCD results.

The three-body parameters obtained through a fit to the lattice spectra will be used to obtain the universal parameters of the $\rho$ and $\omega$ mesons through their pole positions on the second Riemann sheet. The necessary framework is provided through the infinite-volume unitarity (IVU) approach~\cite{Mai:2017vot, Mai:2017bge} corresponding to the above quantization conditions. The part of the scattering amplitude relevant to the emergence of resonance poles is obtained through an integral equation,
%%%%%%%%%%%%%
\begin{align}
    T=\tilde B+\tilde C+\int\frac{d^3l}{(2\pi)^3}\frac{\tilde B+\tilde C}{2E_l(\tilde K^{-1}-\Sigma^{IV})}T\,,
    \label{eq:IVU}
\end{align}
%%%%%%%%%%%%%
where we have suppressed kinematic arguments for brevity; see \cref{eq:QC-tilde} and~\cite{Mai:2021nul,Sadasivan:2021emk,Feng:2024wyg}. Here, $\Sigma^{IV}$ denotes the usual $\rho$ self-energy integral. The corresponding integral equation can be solved through a complex contour deformation in the JLS basis~\cite{Hetherington:1965zza, Sadasivan:2020syi, Feng:2024wyg}, and pole positions extracted as shown in~\cite{Mai:2021nul, Sadasivan:2021emk, Garofalo:2022pux}. The corresponding procedure for finding two-body resonance poles boils down to finding solutions $\{E_2\in\mathds{C}|\tilde K^{-1}=\Sigma^{IV}\}$ on the second Riemann sheet.

The three-body quantization condition is necessarily formulated~\cite{Mai:2021lwb} as an infinitely dimensional determinant equation (PWH). Throughout, this work we assume all momenta up to ${\bm p}_{\rm max}=(2\pi)/L(0,1,1)$ for computational reasons. Related to this is also the fact that for smaller volumes the spectator momentum cutoff leads to negative values of the two-body invariant mass. Since the exact form of the $\tilde K^{-1}$ is not known in this (unphysical) region we cut it off with a simple form factor replacing $\tilde K^{-1}\to (1+e^{-(\sigma-\sigma_0)/M_\pi^2})\tilde K^{-1}, \sigma_0=2M_\pi^2$. 
We have tested other functional forms and values of $\sigma_0$ and ${\bm p}_{\rm max}$, finding no relevant effect on the extracted observables. For further details on cutoff effects in the context of three-body systems see~\cite{Mai:2018djl, Sadasivan:2020syi, Feng:2024wyg}.

\bigskip
%%%%%%%%%%%%%%%%%%%%%%%%%%%%%%%%%%%%
\noindent\emph{Results and discussion}---%
%%%%%%%%%%%%%%%%%%%%%%%%%%%%%%%%%%%%
%For each ensemble, t
The two-body finite-volume spectra consist of three energy eigenvalues located below the first inelastic threshold considering that two pions need to have one unit of momentum for the $T_1^-$. In the three-body spectra, we restrict ourselves to the analysis of the ground states and the first excited state for larger-volume ensembles as shown in \cref{fig:spectra}. While qualitatively also higher levels seem to be predicted by the approach, their quantitative study requires a larger set of lattice operators as well as a formalism update with 
more free parameters due to the close proximity of the next excited state of the $\omega$ meson, the $\omega(1420)$. 

With respect to these data, the method GEN yields best fits with $\chi^2_{\rm d.o.f.}({\rm GEN},\, 305)=1.3,\ \chi^2_{\rm d.o.f.}({\rm GEN},\, 208)=1.6$ including cross-correlations. For more details of the fit results and the obtained parameters, see~\cite{supp}. Global EFT2 fits yield $\chi^2_{\rm d.o.f.}({\rm EFT2},\, 305/208)=3.2$. Provided the precision level of our data, it is noticeable that the EFT2 model is also excessively rigid. Furthermore the corresponding pole positions at the physical point exhibit discrepancies with the empirical data. The EFT4 fit provides a much better description of the finite-volume spectrum $\chi^2_{\rm d.o.f.}({\rm EFT4},\, 208/305)=2.3$ for $g_{\rm EFT4}=5.96(17),\ \delta_{\rm EFT4}=38.8(6.9)\,\MeV,\ M_{V,\rm EFT4}=737(12)\,\MeV,\ a_{\rm EFT4}=0.96(14)\,\GeV^{-1}$, which are, indeed, quite close to the phenomenological values~\cite{Dax:2018rvs, Meissner:1986ka}. We consider EFT4 as our main result, with GEN and EFT2 results providing a measure for systematic uncertainties.

%%%%%%%%%
%%%%%%%%%
\begin{figure}
    \centering
    \includegraphics[width=\linewidth]{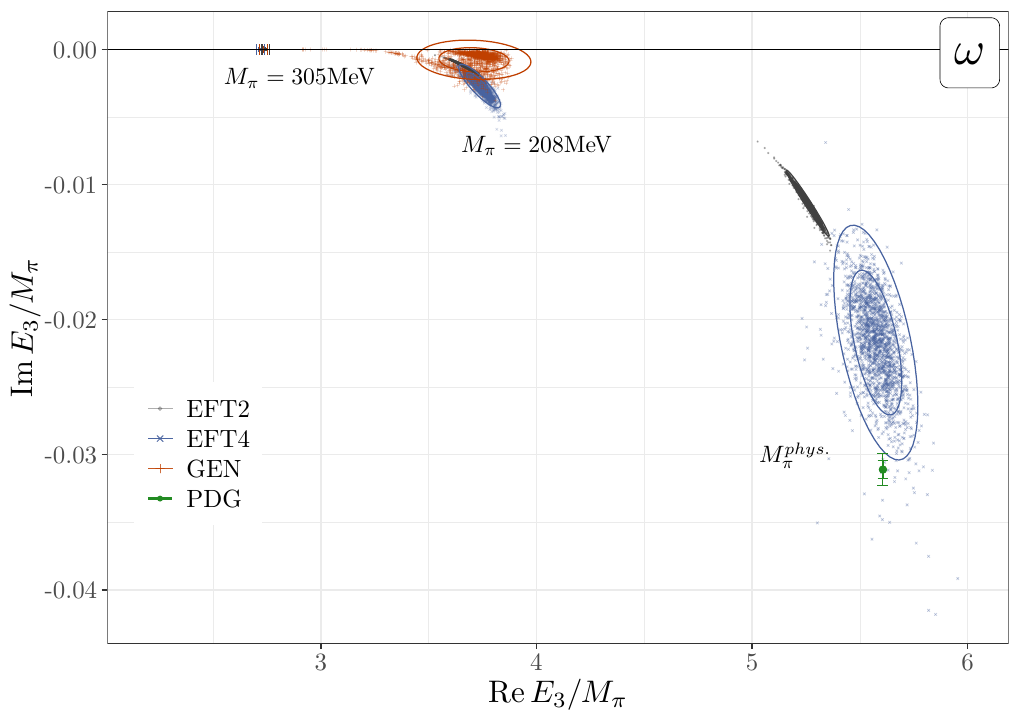}
    \caption{
    Pole positions of the $\omega$ meson at varying pion mass from the IVU approach using generic and effective field theory form of the two- and three-body force. The points are the pole position for each bootstrap sample, whereas the ellipses represent the one- and two-sigma confidence levels. Parameters are fixed by fitting to the available lattice results at $M_\pi=208$ and $M_\pi=305\,\MeV$ allowing one to extrapolate to the physical point. Particle Data Group (PDG) results are quoted by the green error bars for comparison~\cite{ParticleDataGroup:2022pth, CMD-2:2003gqi, Achasov:2003ir}.
    \label{fig:omega}
    }
\end{figure}
%%%%%%%%%
%%%%%%%%%

The resulting $\omega$-pole positions using the IVU formalism~\cref{eq:IVU} with respect to the discussed methods are shown in~\cref{fig:omega} (for $\rho$ see~\cite{supp}). In both cases and for each pion mass, we observe $1\sigma$ agreement between all methods regarding the real part of the pole position, while the imaginary part agrees on the level of $2\sigma$. For the heavier pion mass value, the $\omega$ meson is indeed a bound state (binding energy $\sim80\,\MeV$). We note that the GEN results for the $\omega$ have to be taken with caution since for each pion mass only two volumes are available for the $\omega$ channel with one data point in the relevant energy region, leading possibly to a residual mass and width dependence. Because EFT results connect different pion masses this is not an issue there. The EFT2 pole positions are narrowed to a small region being parametrized only by two parameters $(g,\delta)$. The EFT4 agrees much better with the lattice results on the level of the finite-volume spectra as well as the corresponding GEN pole positions. Extrapolated to the physical pion mass it agrees astonishingly well ($<1\sigma$) with the phenomenological $\rho$- and $\omega$- masses~\cite{ParticleDataGroup:2022pth} and within $2\sigma$ also with their widths. The numerical results in physical units read as
%%%%%
\begin{alignat}{2}
    \sqrt{s_\rho}   
    % &= [5.366(71)-i\,0.455(13)] M_\pi\,, \\
    &= [748.9(10.0)-i\,63.5(1.8)] \, &&\MeV\,, \\
    \sqrt{s_\omega} 
    % &= [5.574(80)-i\,0.022(4) ] M_\pi, \\
    &= [778.0(11.2)-i\,3.0(5) ] \,&&\MeV\,,
\label{eq:main-result}
\end{alignat}
implying for instance also an $\omega-\rho$ mass difference of $29(15)\,\MeV$, which agrees surprisingly well with the result obtained in~\cite{McNeile:2009mx}. The small deviation from the empirical value could be mitigated by taking into account the fact that in some cases too small volumes (see \cref{tab:ens}) could give non-negligible exponential effects and discretization errors.

\bigskip
%%%%%%%%%%%%%%%%%%%%%%%%%%%%%%%%%%%%
\noindent\emph{Conclusions}---%
%%%%%%%%%%%%%%%%%%%%%%%%%%%%%%%%%%%%
We have reported the first-ever lattice QCD estimates of the $\omega$ meson mass and width. One challenge in this calculation consists in particular in the precise estimation of finite-volume spectra from lattice QCD. In resolving the complete low-lying spectrum of states, multihadron operators prove to be essential~\cite{Wilson:2015dqa}. The three-particle operators we use (see~\cite{supp}) require a significant computational effort to compute all the relevant fermion contractions, a task which has only recently become feasible due to advances in algorithms, methods, and computational power. Another challenge concerns the %development of the appropriate formalism 
formalism development to not only map the finite-volume results to the infinite-volume transition amplitudes but to also establish a reliable connection to the pertinent effective field theories, allowing us to perform the so far unprecedented chiral extrapolation of three-body resonance parameters to the physical point. The final results show a good agreement of this theoretical multistep procedure with the empirical values~\cite{ParticleDataGroup:2022pth} regarding the mass of both $\rho$ and $\omega$ mesons. The $\omega$ width turns out slightly smaller than the experimental value, but still agrees within $2\sigma$ uncertainties. We 
%remark that there is 
refer here to an ongoing discussion on the current empirical values~\cite{Hoferichter:2023bjm, Hoferichter:2019mqg, Hoid:2020xjs, Colangelo:2022prz, Colangelo:2018mtw}.

The presented study marks a new milestone in hadron spectroscopy from lattice QCD, paving the way toward understanding more complex systems. For closer to physical pion mass lattice setups the kinematic window to study resonance properties shrinks due to the proximity of the next inelastic thresholds (e.g., $5\pi$). Thus, it may be advantageous to use more stable and widely available results at unphysical pion mass values and extrapolate to the physical point by making use of robust EFT methodology. While standard in two-body studies, such a treatment of resonant three-body systems has not yet been available. Future steps include the assessment of discretization errors as well as the inclusion of larger volumes to reduce the systematic uncertainties further. Applications toward the Roper resonance and $T_{cc}$ are also planned.

\bigskip
%%%%%%%%%%%%%%%%%%%%%%%%%%%%%%%%%%%%
\noindent\emph{Acknowledgments}---%
%%%%%%%%%%%%%%%%%%%%%%%%%%%%%%%%%%%%
We thank the CLQCD Collaborations for providing us the gauge configurations with dynamical fermions~\cite{Hu2024}, which are generated on the HPC Cluster of ITP-CAS, the Southern Nuclear Science Computing Center(SNSC), the Siyuan-1 cluster supported by the Center for High Performance Computing at Shanghai Jiao Tong University, and the Dongjiang Yuan Intelligent Computing Center.

H.Y. is grateful to Y.~Meng, P.~Sun, J.~Wu, and other members of CLQCD for helpful discussions. H.Y. also thanks C.~Thomas and D.~Wilson for valuable questions and discussions. Software \verb|QUDA|~\cite{Clark2010, Babich2011, Clark2016} is used to solve the perambulators, as performed in~\cite{Yan2024b}. H.Y., C.L., and L.L. acknowledge support from NSFC under Grants No.~12293060, No.~12293061, No.~12293063, No.~12175279, and No.~11935017. L.L. is thankful for support from the Strategic Priority Research Program of the Chinese Academy of Sciences with Grant No.~XDB34030301 and the Guangdong Major Project of Basic and Applied Basic Research No. 2020B0301030008. This project was funded by the Deutsche Forschungsgemeinschaft (DFG,
German Research Foundation) as part of the CRC 1639 NuMeriQS – Project No.~511713970 and by the MKW NRW under the funding code No.~NW21-024-A. M.M. thanks M. Döring, D. Sadasivan, H. Akdag, and P.~C.~Bruns for useful discussions. 
This work is supported by the Deutsche Forschungsgemeinschaft (DFG, German Research Foundation) through the Sino-German Collaborative Research Center CRC110 “Symmetries and the Emergence of Structure in QCD” (DFG Project ID No.~196253076 - TRR 110). 
The work of M.M. was further funded through the Heisenberg Programme by the Deutsche Forschungsgemeinschaft (DFG, German Research Foundation) – 532635001, and the National Science Foundation Grant No. PHY-2012289.
The work of U.G.M. was supported in part by the CAS President’s International Fellowship Initiative (PIFI) (Grant No. 2025PD0022). Part of the simulations were performed on the High-performance Computing Platform of Peking University, and the QBiG GPU cluster at HISKP/Univ. Bonn.
%%%%%%%%%%%%%%%%%%%%%%%%%%%%%%%%%%%%

% \onecolumngrid
%%%%%%%%%%%%%%%%%%%%%%%%%%%%%%%%%%%%
% \bibliographystyle{apsrev4-2}
\bibliography{paper}
%%%%%%%%%%%%%%%%%%%%%%%%%%%%%%%%%%%%

\clearpage
\appendix
\onecolumngrid
\setcounter{equation}{0}
\setcounter{figure}{0}
\setcounter{table}{0}
\makeatletter
\renewcommand{\theequation}{S\arabic{equation}}
\renewcommand{\thefigure}{S\arabic{figure}}
\renewcommand{\thetable}{S\arabic{table}}
\setcounter{secnumdepth}{2}

%%%%%%%%%%%%%%%%%%%%%%%%%%%%%%%%%%%%
\section*{Supplemental Material}
%%%%%%%%%%%%%%%%%%%%%%%%%%%%%%%%%%%%

%%%%%%%%%%%%%%%%%%%%%%%%%%%%%%%%%%%%
\subsection{Table of operators}
\label{SUPP/SEC:operators}
%%%%%%%%%%%%%%%%%%%%%%%%%%%%%%%%%%%%
Each operator used in this work is a linear combination of quark bilinear or the product of bilinears. First, we construct the operators in the flavor space. Let $\rho^+_i/\pi^+ = \bar{d} \gamma_{i/5} u, [\rho/\pi]^u = \bar{u} \gamma_{i/5} d$, \textit{etc}. For $I=1$ $\pi\pi$ channel, the following flavor structure is used:
\begin{equation}
\begin{cases}
| \rho \rangle &= -\rho^+, \\
| \pi\pi \rangle &= \frac{1}{2} \left[ -\pi^+ \pi^u + \pi^+ \pi^d + \pi^u \pi^+ - \pi^d \pi^+ \right],
\end{cases} \\
\end{equation}
where $| \rho \rangle$ and $| \pi\pi \rangle$ are one-body and two-body operators, respectively.

For $I=0$ $\pi\pi\pi$ channel, the following structure of bilinears is designed to enforce the isoscalar projection
\begin{equation}
\begin{cases}
| \omega \rangle &= \frac{1}{\sqrt{2}} (\omega^u + \omega^d), \\
| \rho\pi \rangle &= -\frac{1}{\sqrt{3}} \left[ \rho^+ \pi^- + \rho^- \pi^+ + \frac{1}{2} [\rho^u \pi^u - \rho^u \pi^d - \rho^d \pi^u + \rho^d \pi^d] \right], \\
| \pi\pi\pi \rangle &= \frac{1}{\sqrt{12}} \left[ -\pi^+\pi^u\pi^- + \pi^+\pi^d\pi^- + \pi^u\pi^+\pi^- - \pi^d\pi^+\pi^- + \pi^+\pi^-\pi^u -\pi^+\pi^-\pi^d \right. \\
& \qquad \left. -\pi^-\pi^+\pi^u + \pi^-\pi^+\pi^d - \pi^u\pi^-\pi^+ + \pi^d\pi^-\pi^+ + \pi^-\pi^u\pi^+ -\pi^-\pi^d\pi^+ \right],
\end{cases}
\end{equation}
The operator set includes all one-, two-, and three-body operators to overlap with the dynamical channels in this study.

Second, the two-body operators are further projected onto the $T_1^-$ irrep in the momentum space to detect the P-wave. With \verb|OpTion|~\cite{Yan2024}, the explicit forms of the operators are printed. For convenience, the operators are represented as
\begin{equation}
\begin{cases}
    O_{\text{one}} = \sum_{i} \eta_i \omega_{\mu_i}(0), \\
    O_{\text{two}} = \sum_{i} \eta_i \rho_{\mu_i}(\vec{p}_i) \pi(-\vec{p}_i), \\
    O_{\text{three}} = \sum_{i} \eta_i \pi_{\mu_i}(\vec{p}_{i1}) \pi_{\mu_i}(\vec{p}_{i2}) \pi(-\vec{p}_{i1}-\vec{p}_{i2}),
\end{cases}
\end{equation}
and are uniquely identified by the parameters $\eta_i$, $\mu_i$ and $\vec{p}_i$. $\mu_i \in \{ 0, 5, x, y, z \}$ is the Cartesian gamma matrices and corresponds to $\gamma_0, \gamma_5, \gamma_x, \gamma_y, \gamma_z$. $\vec{p}_i$ is the momenta for each meson. For convenience, the parameters are denoted by
\begin{equation}
\eta_{\mu_i}^{\alpha_{i1}(;\alpha_{i2})},
\end{equation}
where $\alpha_i$ one-to-one corresponds to the momentum vector. One direction notation in $\alpha_i$ means one unit of momentum in that direction. For example, $\alpha = yz \to \vec{p} = [011]$, $\alpha = -2x \to \vec{p} = [-200]$, \textit{etc}. The space projection coefficients are shown in Tab.~\ref{tab:operators}.

%%%%%%%%%%%%%%%%%%%%
%%%%%%%%%%%%%%%%%%%%
\begin{table*}[htbp]
\addtolength{\tabcolsep}{6pt}
\begin{tabular}{ccc}
\toprule
channel & type & operator \\
\midrule
\multirow{3}{*}{$\pi\pi$} & \multirow{1}{*}{one} & $(+1)^{0}_{z}$ \\
\cmidrule(lr){2-3}
& \multirow{2}{*}{two} & $(+1)^{0}_{5}$ \\
& & $(+1)^{x}_{5}, (+1)^{-x}_{5}, (+1)^{y}_{5}, (+1)^{-y}_{5}, (+1)^{z}_{5}, (+1)^{-z}_{5}$ \\
\midrule
\multirow{4}{*}{$\pi\pi\pi$} & \multirow{1}{*}{one} & $(+1)^{0}_{z}$ \\
\cmidrule(lr){2-3}
& \multirow{2}{*}{two} & $(-1)^{-y}_{x}, (+1)^{y}_{x}, (+1)^{-x}_{y}, (-1)^{x}_{y}$ \\
& & $(-1)^{xy}_{x}, (+1)^{xy}_{y}, (+1)^{x,-y}_{x}, (+1)^{x,-y}_{y}, (-1)^{-x,y}_{x}, (-1)^{-x,y}_{y}, (+1)^{-x,-y}_{x}, (-1)^{-x,-y}_{y}$ \\
\cmidrule(lr){2-3}
& \multirow{1}{*}{three} & $(+1)^{xy;-y}_{5}, (-1)^{x,-y;y}_{5}, (-1)^{-x,y;-y}_{5}, (+1)^{-x,-y;y}_{5}, (-1)^{xy;-x}_{5}, (+1)^{-x,y;x}_{5}, (+1)^{x,-y;-x}_{5}, (-1)^{-x,-y;x}_{5}$ \\
\bottomrule
\end{tabular}
\caption{Operators used to interpolate the $\pi\pi$ and $\pi\pi\pi$ system. Each irrep contains both one-meson type, two-meson, and possible three-meson type operators. The symbols $\eta_{\mu_i}^{\alpha_{i1}(;\alpha_{i2})}$ are defined in the context. The overall constants are ignored.}
\addtolength{\tabcolsep}{-6pt}
\label{tab:operators}
\end{table*}
%%%%%%%%%%%%%%%%%%%%
%%%%%%%%%%%%%%%%%%%%

Non-local operators with covariant derivatives are tested. The resulting spectra do not lead to noticeable change and as a result, this kind of operators are disposed out of the operator set.

%%%%%%%%%%%%%%%%%%%%%%%%%%%%%%%%%%%%
\subsection{Topologies of the contraction diagrams}
\label{SUPP/SEC:topologies}
\begin{figure}[htbp]
\centering
\raisebox{-0.5\height}{\includegraphics[width=0.12\columnwidth]{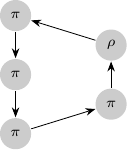}}
\quad
\raisebox{-0.5\height}{\includegraphics[width=0.138\columnwidth]{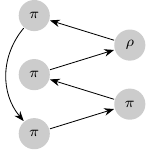}}
\quad
\raisebox{-0.5\height}{\includegraphics[width=0.12\columnwidth]{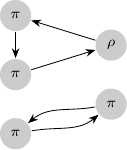}}
\quad
\raisebox{-0.5\height}{\includegraphics[width=0.12\columnwidth]{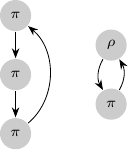}}
\quad
\raisebox{-0.5\height}{\includegraphics[width=0.12\columnwidth]{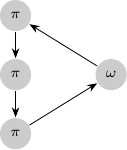}}
\quad
\raisebox{-0.5\height}{\includegraphics[width=0.12\columnwidth]{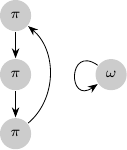}}
\caption{Topologies of the diagrams for $\pi\pi\pi \to \rho\pi$ and $\pi\pi\pi \to \omega$.}
\label{fig:topologies-32-31}
\end{figure}

%%%%%%%%%%%%%%%%%%%%%%%%%%%%%%%%%%%%
After projecting to isospin $I=0$, only certain topologies remain. The topologies of $\pi\pi\pi \to \pi\pi\pi$ are illustrated in the main text. The schematics of other diagrams including the three-pion operators, \textit{i.e.}, $\pi\pi\pi \to \rho\pi$ and $\pi\pi\pi \to \omega$, are shown in Fig.~\ref{fig:topologies-32-31}.

%%%%%%%%%%%%%%%%%%%%
%%%%%%%%%%%%%%%%%%%%
%%%%%%%%%%%%%%%%%%%%
%%%%%%%%%%%%%%%%%%%%

\begin{figure}[htbp]
\centering
\raisebox{-0.5\height}{\includegraphics[width=0.12\columnwidth]{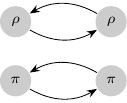}}
\quad
\raisebox{-0.5\height}{\includegraphics[width=0.12\columnwidth]{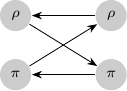}}
\quad
\raisebox{-0.5\height}{\includegraphics[width=0.12\columnwidth]{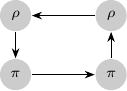}}
\quad
\raisebox{-0.5\height}{\includegraphics[width=0.12\columnwidth]{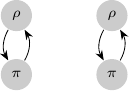}}
\quad
\raisebox{-0.5\height}{\includegraphics[width=0.12\columnwidth]{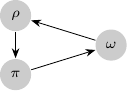}}
\quad
\raisebox{-0.5\height}{\includegraphics[width=0.12\columnwidth]{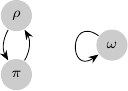}}
\raisebox{-0.5\height}{\includegraphics[width=0.12\columnwidth]{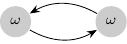}}
\caption{Topologies of the diagrams for $\rho\pi \to \rho\pi$, $\rho\pi \to \omega$, and $\omega \to \omega$.}
\label{fig:topologies-22-21-11}
\end{figure}

Topologies in the two-pion and one-pion sectors are shown in Fig~\ref{fig:topologies-22-21-11}. The type of topologies is the same as those for $I=0$ $\pi\pi$ scattering, with one of the $\pi$ replaced by the $\rho$ meson and the target resonance $\sigma$ replaced by the $\omega$.

%%%%%%%%%%%%%%%%%%%%
%%%%%%%%%%%%%%%%%%%%
%%%%%%%%%%%%%%%%%%%%
%%%%%%%%%%%%%%%%%%%%

%%%%%%%%%%%%%%%%%%%%%%%%%%%%%%%%%%%%
\subsection{Spectra}
\label{SUPP/SEC:spectra}
%%%%%%%%%%%%%%%%%%%%%%%%%%%%%%%%%%%%

In this section, we present the details of extracting finite-volume energy levels from the correlation matrices $C_{ij}(t)$. To determine the lowest levels, we solve the GEVP for each ensemble,
\begin{equation}
C(t) v_n(t, t_0) = \lambda_n(t, t_0) C(t_0) v_n(t, t_0)\,,
\end{equation}
where $\lambda_n(t, t_0)$ are the eigenvalues corresponding to the eigenvectors $v_n(t, t_0)$. For the $\pi\pi$ channel, we use $3 \times 3$ matrices and for the $\pi\pi\pi$ channel, we use $4 \times 4$ matrices using the operators in \cref{tab:operators}. The choice of the reference time $t_0$ has a negligible impact on the resulting spectra.

\begin{figure}[htbp]
\centering
\includegraphics[width=0.48\columnwidth]{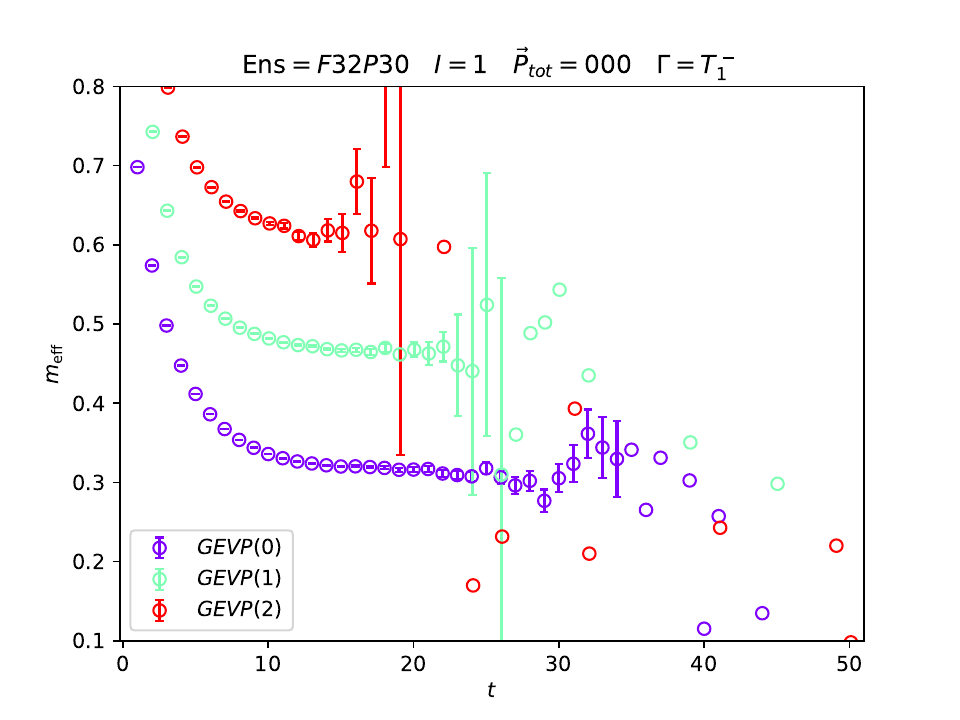}
\includegraphics[width=0.48\columnwidth]{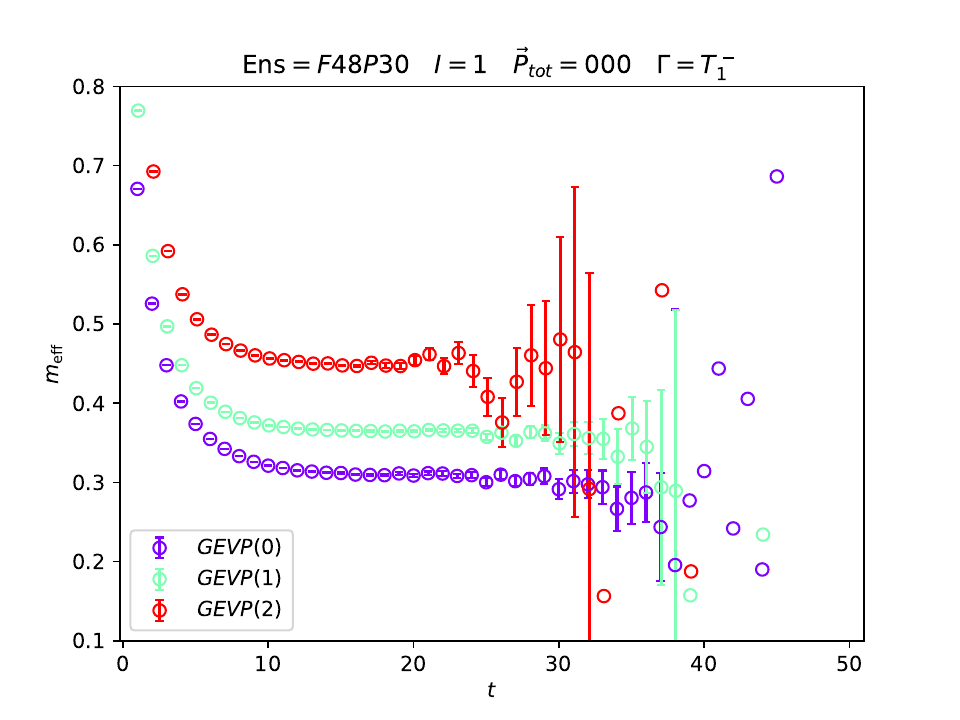}
\includegraphics[width=0.48\columnwidth]{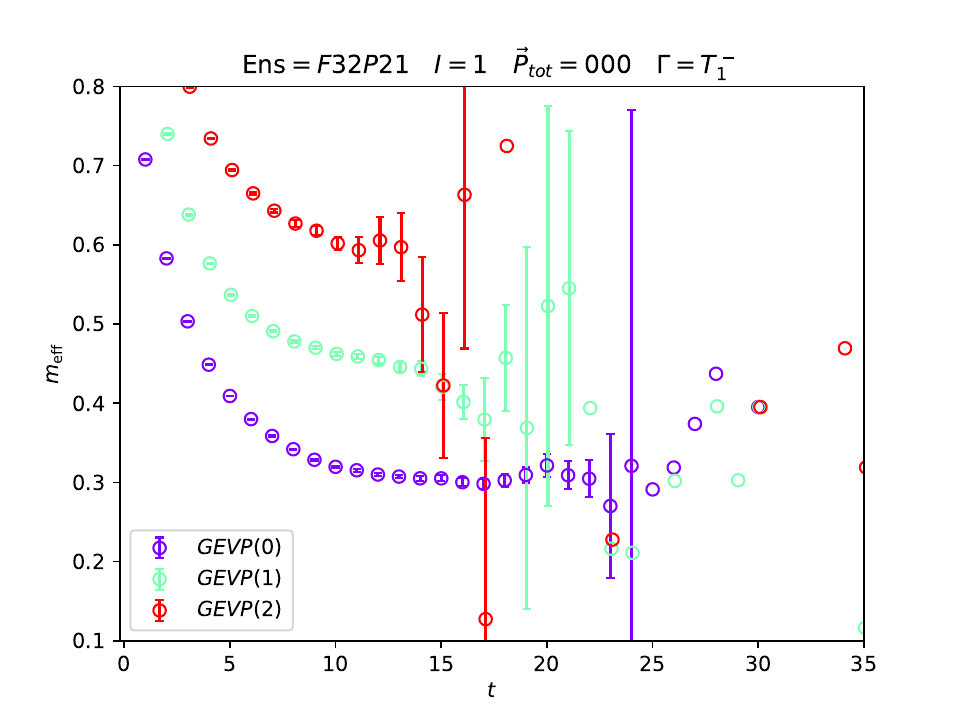}
\includegraphics[width=0.48\columnwidth]{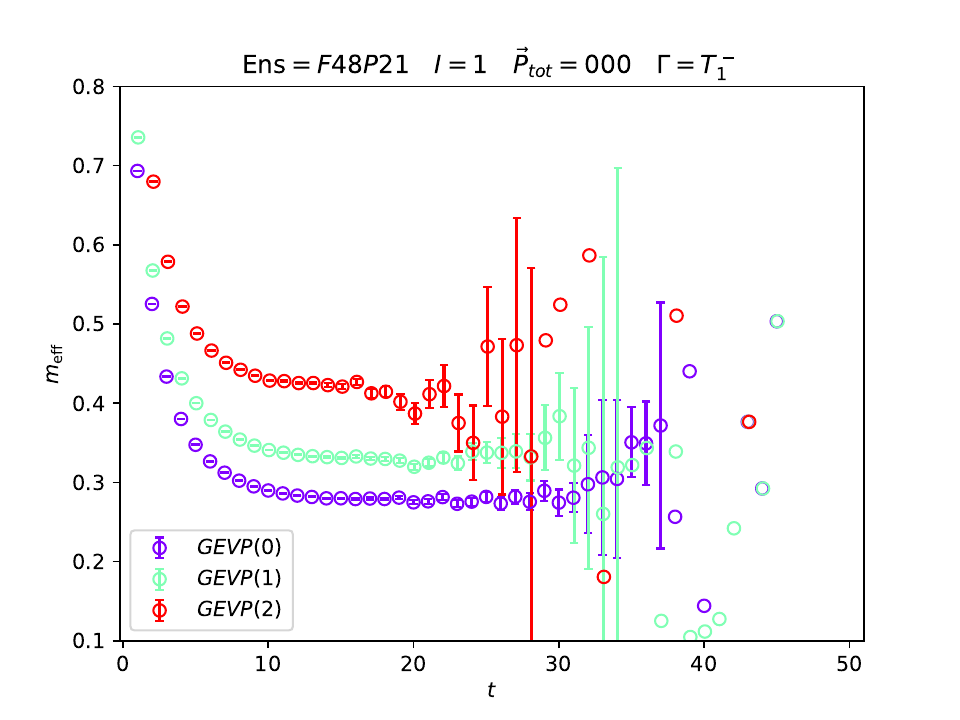}
\caption{Effective masses of the GEVP eigenvalues in the $I=1 \, \pi\pi$ channel.}
\label{fig:pipi-I=1-meff}
\end{figure}

The effective mass of the eigenvalues in the $\pi\pi$ and $\pi\pi\pi$ channel are shown in Figs.~\ref{fig:pipi-I=1-meff} and \ref{fig:pipi-I=1-meff}, respectively. The plots exhibit plateaus at large $t$. The thermal pollution in the $\pi\pi$ channel is eliminated by shifting the correlation matrices before solving GEVP. The thermal pollution in the $\pi\pi\pi$ channel is not evident with the current precision, and is, thus, ignored in the analysis.

%%%%%%%%%%%%%%%%%%%%
%%%%%%%%%%%%%%%%%%%%

\begin{figure}[htbp]
\centering
\includegraphics[width=0.48\columnwidth]{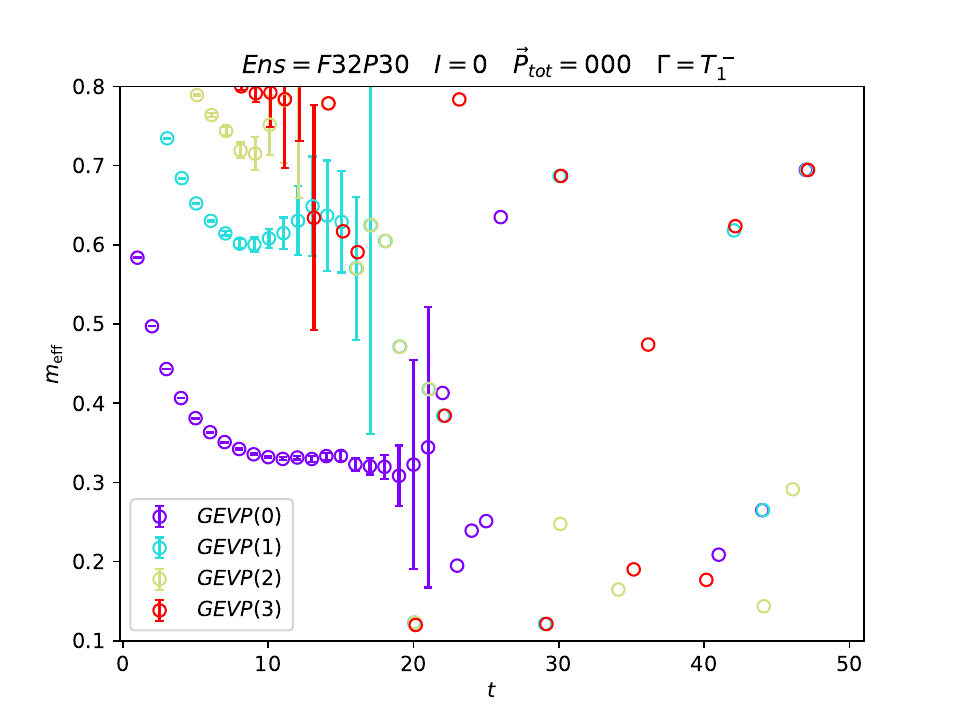}
\includegraphics[width=0.48\columnwidth]{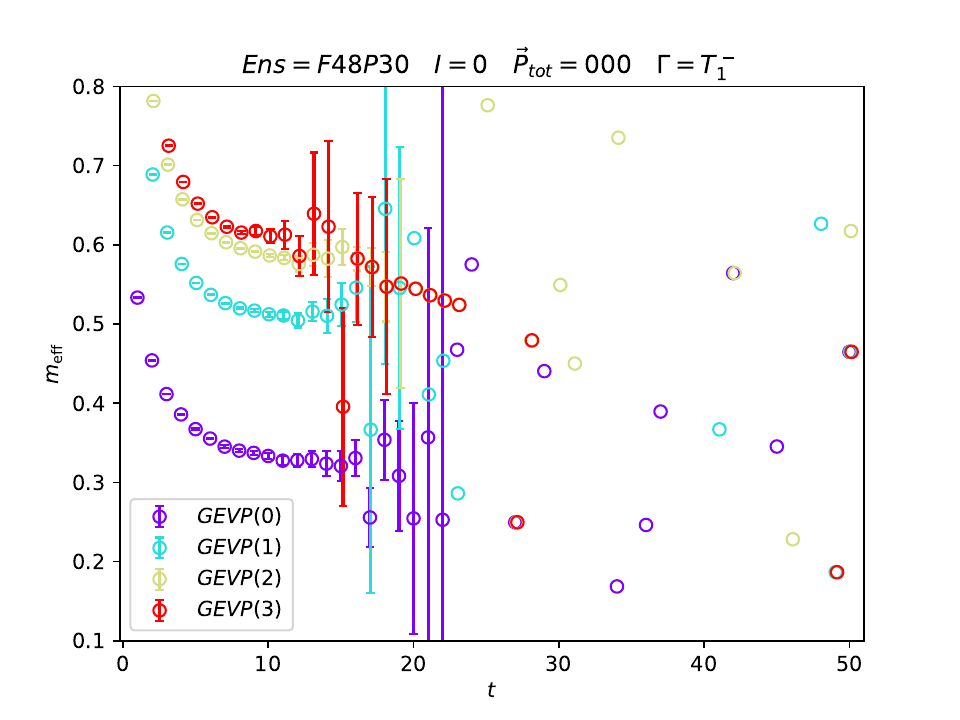}
\includegraphics[width=0.48\columnwidth]{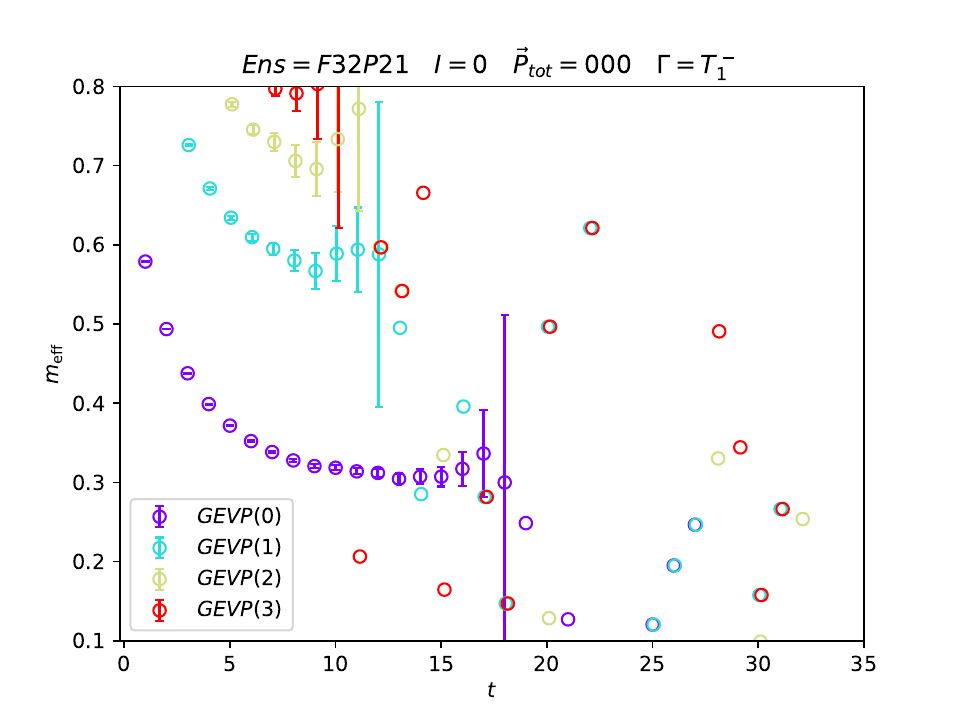}
\includegraphics[width=0.48\columnwidth]{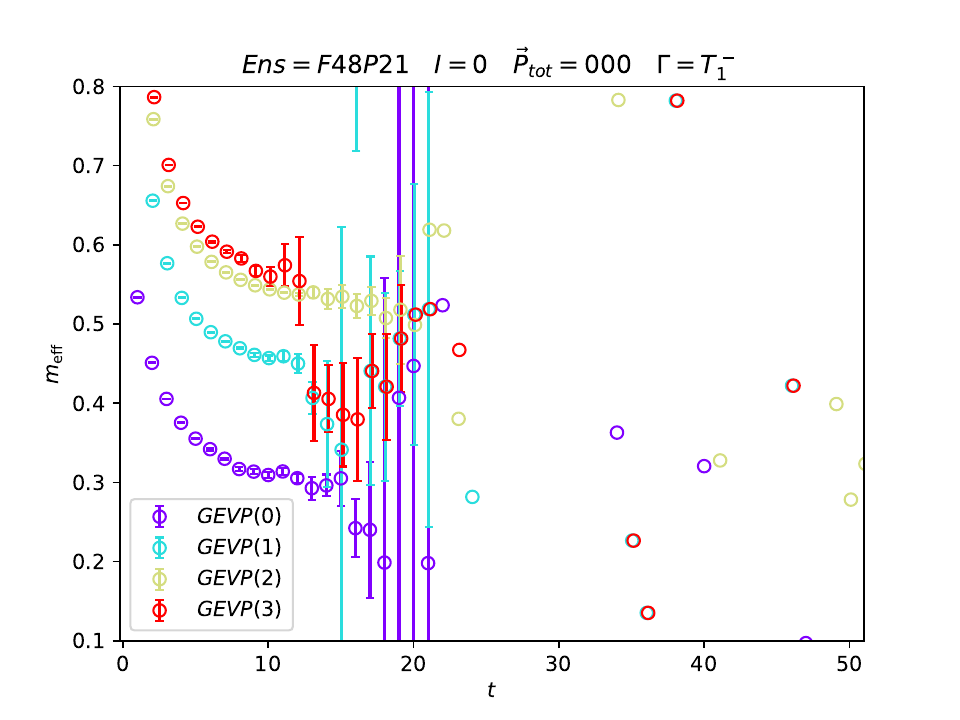}
\caption{Effective masses of the GEVP eigenvalues in the $I=0 \, \pi\pi\pi$ channel.}
\label{fig:pipipi-I=0-meff}
\end{figure}
%%%%%%%%%%%%%%%%%%%%
%%%%%%%%%%%%%%%%%%%%

The energy levels of the $n^{\text{th}}$ excited states are extracted by doing a two-state fit of $\lambda_n(t, t_0)$:
\begin{equation}
\lambda_n(t, t_0) = (1-A_n) \operatorname{e}^{-E_n(t-t_0)} + A_n \operatorname{e}^{-E_n^{\prime}(t-t_0)},
\label{eq:2state_fit}
\end{equation}
where $E_n$ is the $n^{\text{th}}$ energy level.

\begin{figure}[htbp]
\centering
\includegraphics[width=0.6\columnwidth]{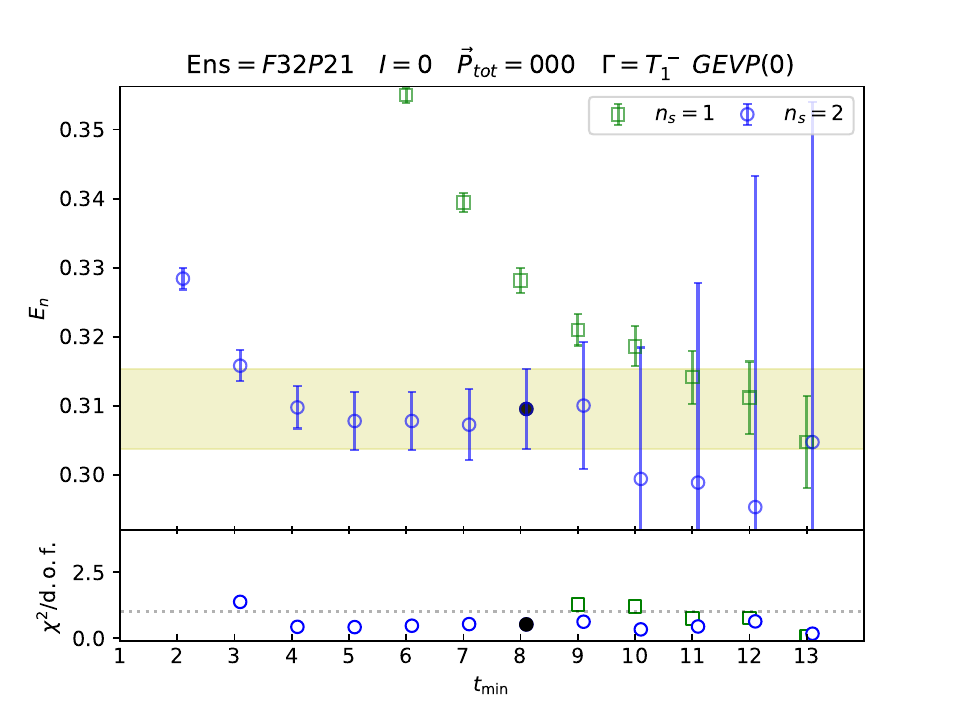}
\caption{Stability plot for the ground state fit in the $I=0 \, \pi\pi\pi$ channel of ensemble F32P21. The blue circles are the result of the two-state fit
\ref{eq:2state_fit}, while the green squares represent the results from a 
one-state fit given by $\lambda_n(t, t_0) =  \operatorname{e}^{-E_n(t-t_0)}$. 
The chosen value of $t_{\mathrm{min}}$ in the two-state fit is represented as a filled point.}
\label{fig:stability}
\end{figure}

We inspected the dependence of the levels on the fitting range and chose the starting point $t_{\mathrm{min}}$ such that the fit yields a reasonable $\chi^2_{\rm dof}$ and remains stable against changes in the fitting range. An example of this stability analysis for the ground state in the $\pi\pi\pi$ channel of ensemble F32P21 is shown in \cref{fig:stability}. Statistical uncertainties are estimated from $2000$ Bootstrap samples.

%%%%%%%%%%%%%%%%%%%%
%%%%%%%%%%%%%%%%%%%%
%%%%%%%%%%%%%%%%%%%%
%%%%%%%%%%%%%%%%%%%%

% \clearpage
%%%%%%%%%%%%%%%%%%%%%%%%%%%%%%%%%%%%
\subsection{Fit results}
\label{SUPP:tab_fits}
%%%%%%%%%%%%%%%%%%%%%%%%%%%%%%%%%%%%

\begin{figure}[htbp]
\centering
\includegraphics[width=0.49\columnwidth]{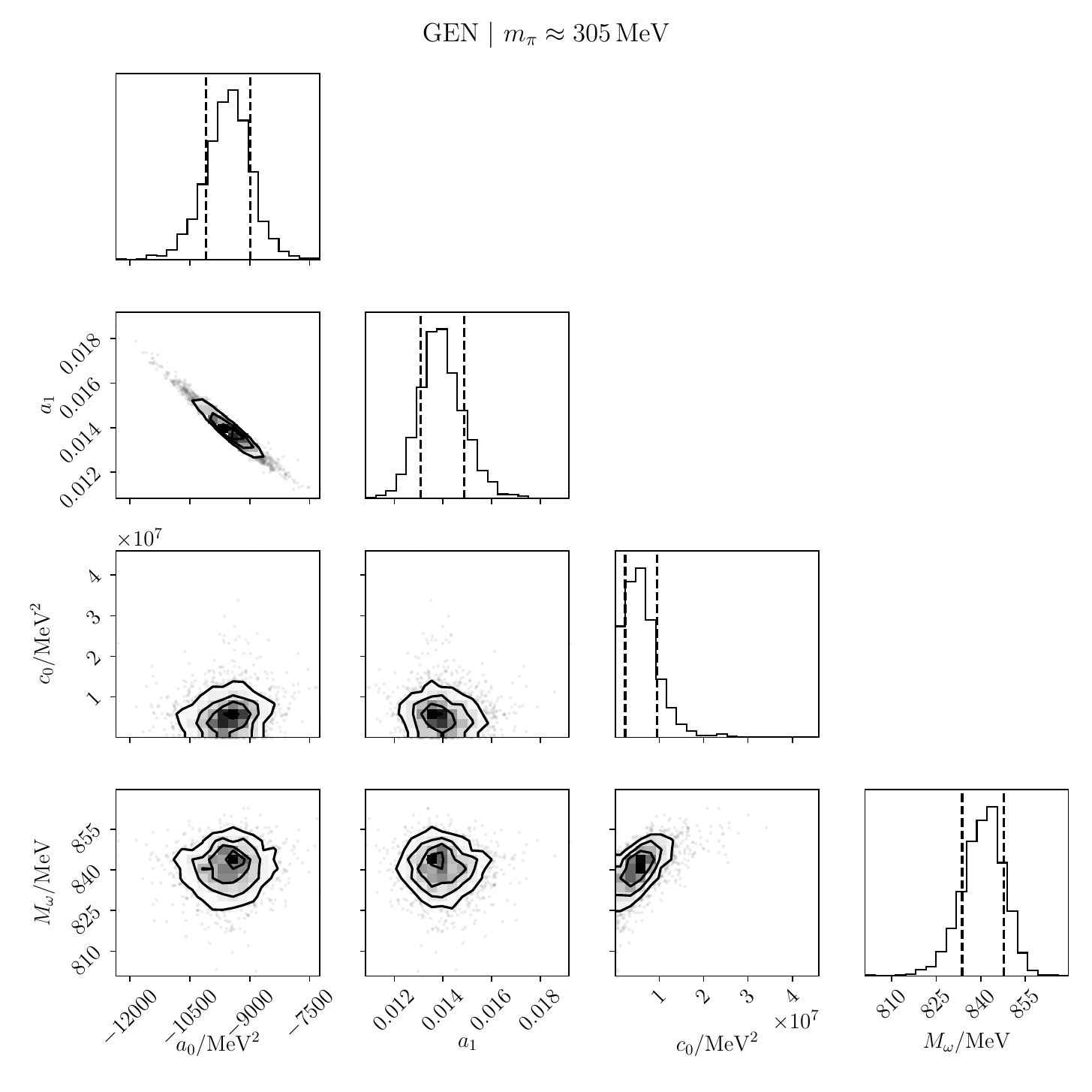}
\hfill
\includegraphics[width=0.49\columnwidth]{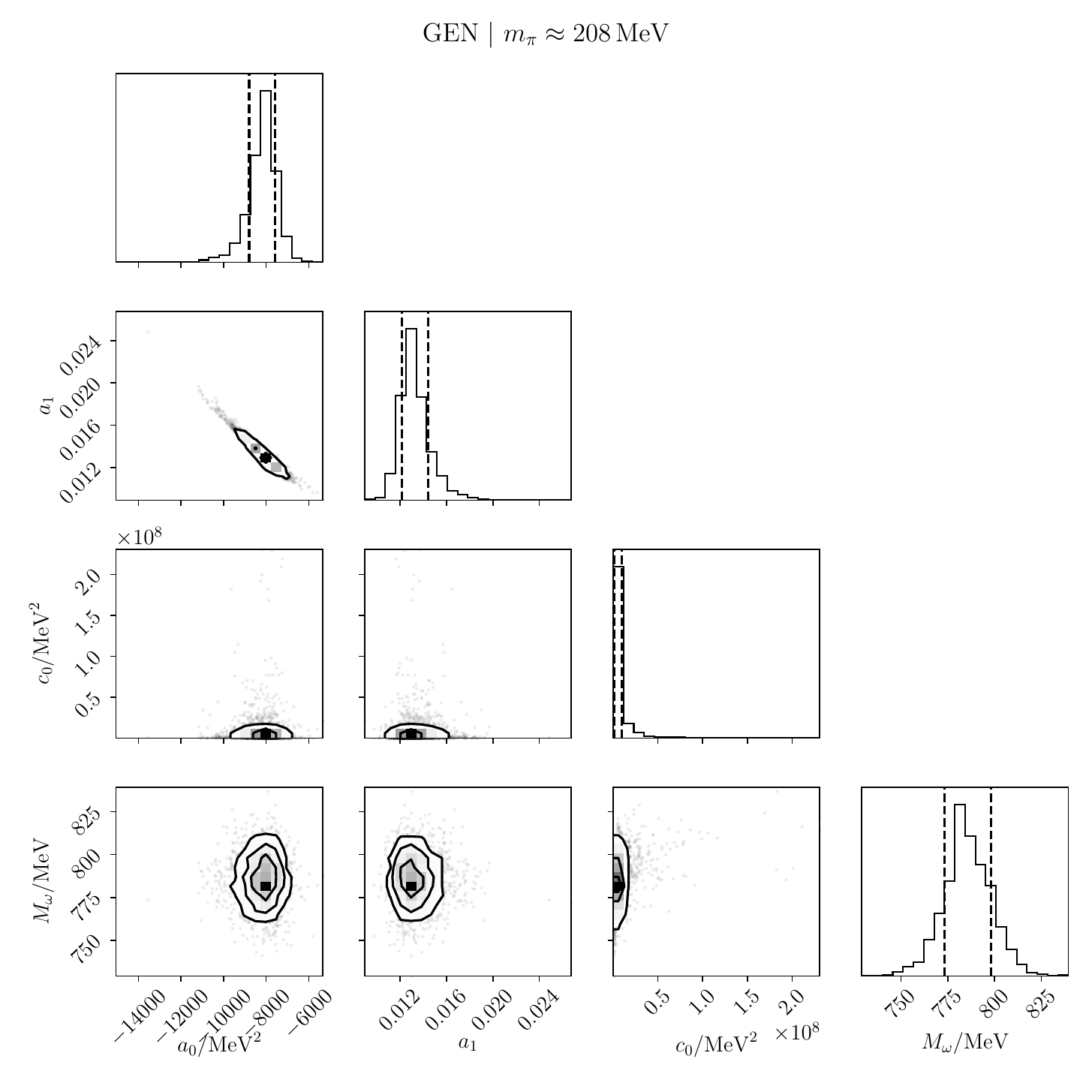}
\caption{Corner plots for GEN parametrizations with respect to fits at heavy (left panel) and light (right panel) pion masses.}
\label{SUPP/fig:GENcorner}
\end{figure}

\begin{figure}[htbp]
\centering
\includegraphics[width=0.30\columnwidth]{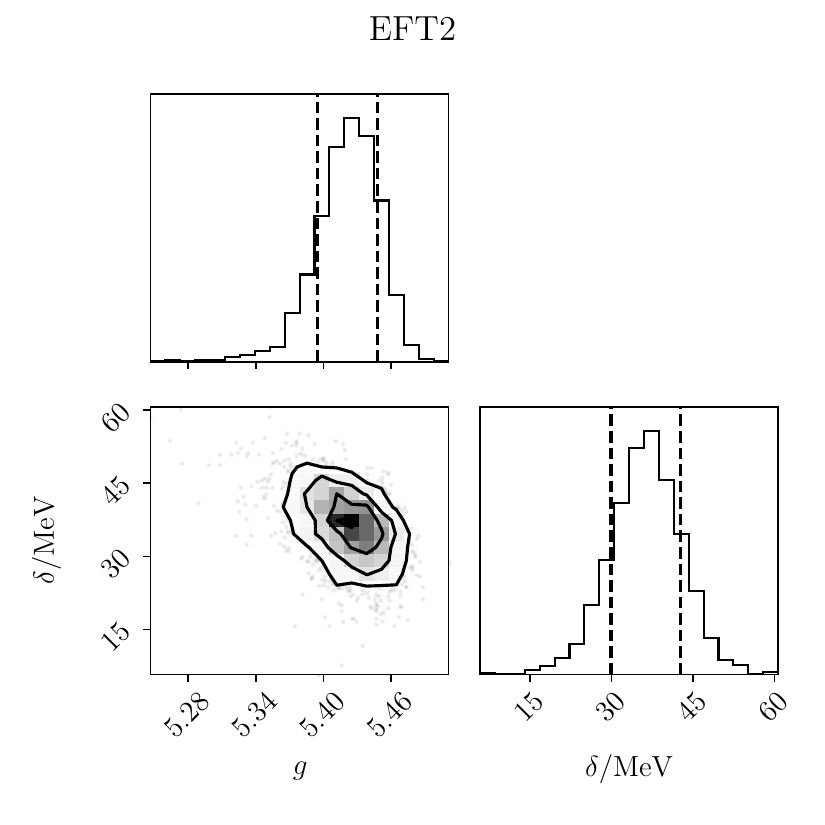}
\hfill
\includegraphics[width=0.6\columnwidth]{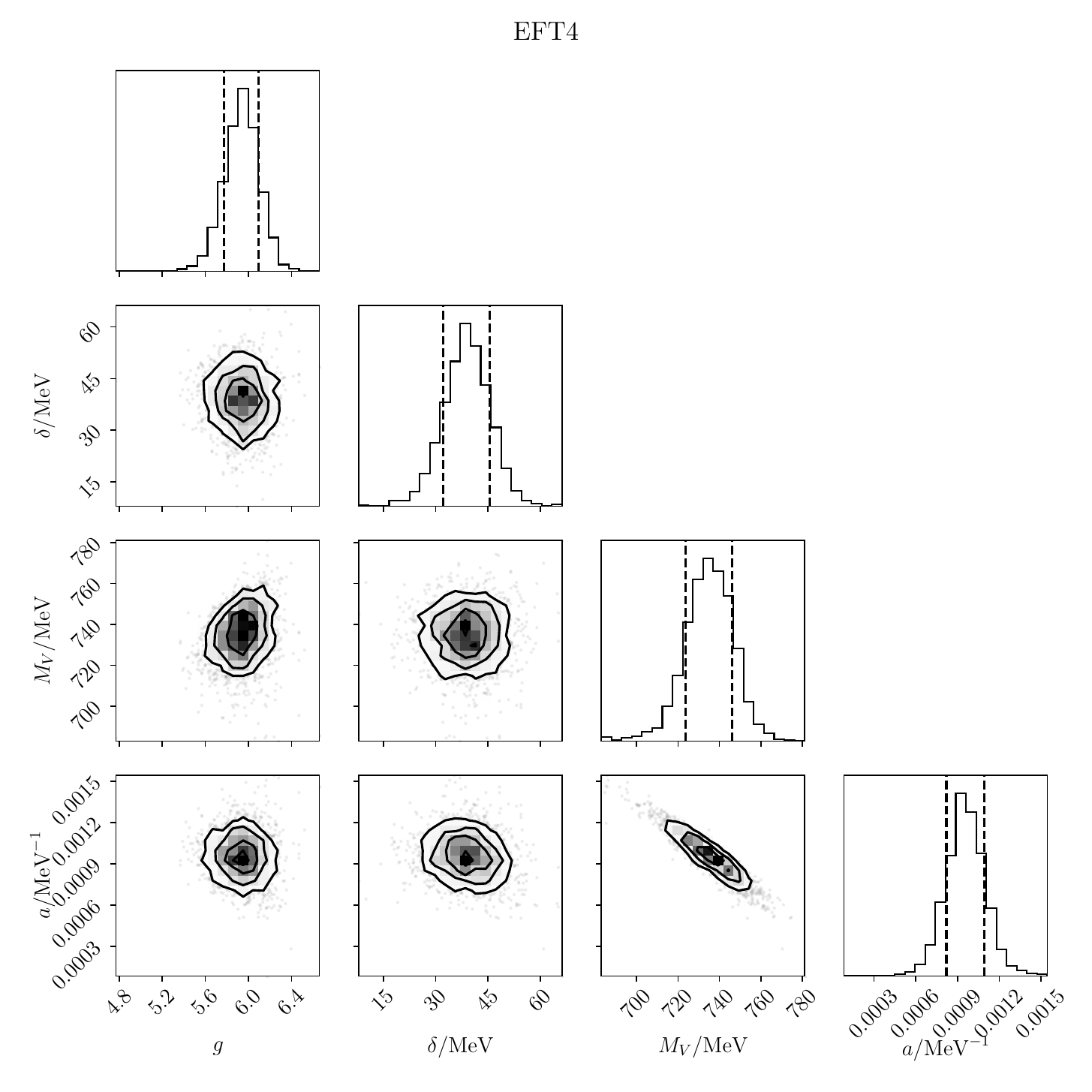}
\caption{Corner plots for the EFT2 (left panel) and EFT4 (right panel) parametrizations with respect to simultaneous fits to light and heavy pion mass ensembles.}
\label{SUPP/fig:EFTcorner}
\end{figure}

Details of the fit results for the GEN(305), GEN(208), EFT2(208/305) and EFT4(208/305) methods of parametrizing two- and three-body force. All values are estimated for the ${\bm p}_{\rm max}=(2\pi)/L(0,1,1)$ spectator momentum cutoff and left-hand cutoff form-factor $\tilde K^{-1}\to (1+e^{-(\sigma-\sigma_0)/M_\pi^2})\tilde K^{-1}$. Statistical uncertainties are estimated from $2000$ Bootstrap samples. For the EFT4 fit, we also estimated the systematic uncertainty from the exponential finite-volume effect by excluding energy levels from the F32P21 ensemble, which has the smallest $M_\pi L$. The difference in the fit parameters with and without this ensemble is reported as the second error. This uncertainty remains within or slightly above the statistical error, suggesting that the finite-volume effect does not significantly impact our fit. The distribution and covariances of the parameters of GEN and EFT fits are shown in the corner plots~\cref{SUPP/fig:GENcorner} and~\cref{SUPP/fig:EFTcorner}, respectively.

%%%%%%%%%%%%%%%%%%%%
%%%%%%%%%%%%%%%%%%%%
\begin{table}[h]
\centering
\footnotesize
\renewcommand{\arraystretch}{1.5}
\addtolength{\tabcolsep}{4pt}
\begin{tabular}{c|cccl}
\toprule
$M_{\pi}/\MeV$ & $\tilde{K}^{-1}$ & $c_{11}$ & $\chi^2_{\rm dof}$ & Parameters \\
\midrule
\multirow{1}{*}{305} & $a_0 + a_1 \sigma$ & $\frac{c_0}{s - M_\omega^2}$ & $1.25$ & $a_0 = -9556(597) \,\mathrm{MeV}^2,\ a_1 = 0.01393(94),\ c_0 = 5.7(4.2) \times 10^6 \,\mathrm{MeV}^2,\ M_{\omega} = 841.5(7.3) \,\mathrm{MeV}$ \\
\midrule
\multirow{1}{*}{208} & $a_0 + a_1 \sigma$ & $\frac{c_0}{s - M_\omega^2}$ & $1.57$ & $a_0 = -8068(718) \,\mathrm{MeV}^2,\ a_1 = 0.0130(14),\ c_0 = 2(17) \times 10^6 \,\mathrm{MeV}^2,\ M_{\omega} = 784(13) \,\mathrm{MeV}$ \\
\midrule
\multirow{2}{*}{305/208} & $\frac{\sigma - M_{\rho}^2}{2 g^2}$ & $c_{11}^{\mathrm{EFT}}$ & $3.16$ & $g = 5.432(29),\ \delta = 36.2(6.7) \ \mathrm{MeV}$ \\
& $\frac{\sigma - M_{\rho}^2(M_V, a)}{2 g^2}$ & $c_{11}^{\mathrm{EFT}}$ & $2.29$ & $g = 5.960(167)[5],\ \delta = 38.8(6.9)[7.3] \ \mathrm{MeV},\ M_V = 737(12)[2] \ \mathrm{MeV},\ a = 0.00096(14)[4] \ \mathrm{MeV}^{-1}$ \\
\bottomrule
\end{tabular}
\caption{Results for simultaneous 2/3-body fit.}
\addtolength{\tabcolsep}{-6pt}
\label{tab:23fit}
\end{table}
%%%%%%%%%%%%%%%%%%%%
%%%%%%%%%%%%%%%%%%%%

%%%%%%%%%%%%%%%%%%%%
%%%%%%%%%%%%%%%%%%%%
%%%%%%%%%%%%%%%%%%%%
%%%%%%%%%%%%%%%%%%%%

%%%%%%%%%%%%%%%%%%%%
%%%%%%%%%%%%%%%%%%%%
%%%%%%%%%%%%%%%%%%%%
%%%%%%%%%%%%%%%%%%%%

\begin{figure}
    \centering
    \includegraphics[width=0.6\linewidth]{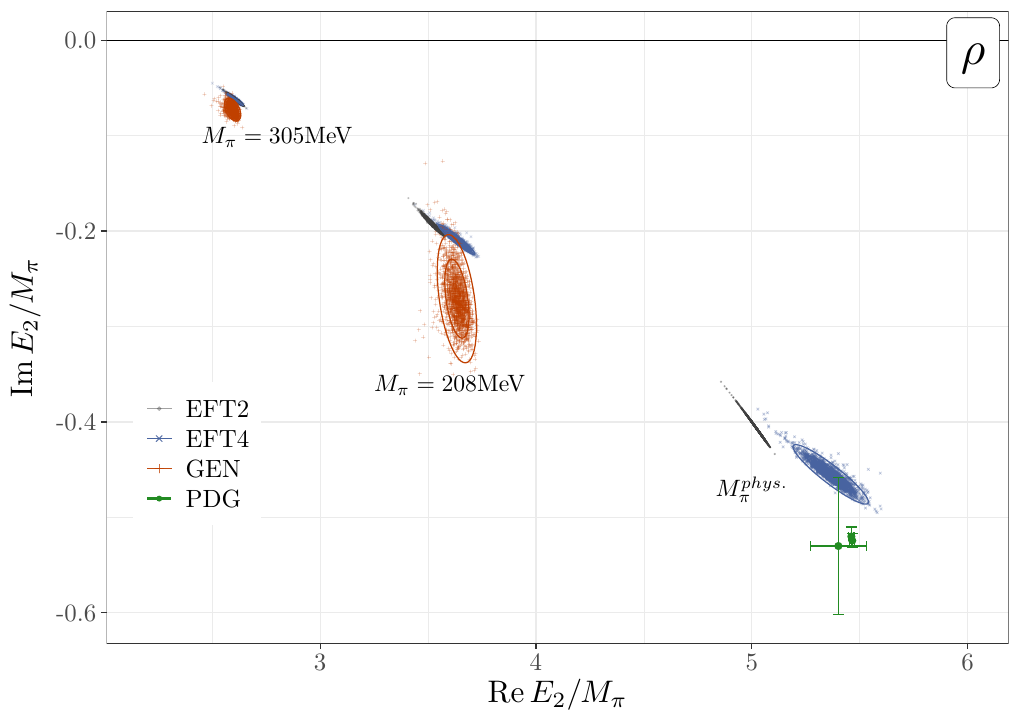}\\
    \caption{
    Pole positions of the $\rho$-meson at varying pion mass from the IVU approach using generic and effective field theory form of the two- and three-body force, as described in \cref{fig:omega} in the main text. PDG results are quoted by the green error bars for comparison~\cite{ParticleDataGroup:2022pth, Garcia-Martin:2011nna, Pelaez:2004xp, Colangelo:2001df}.
    \label{fig:rho}
    }
\end{figure}

The resulting $\rho$ pole positions using the procedure described in Sec~\emph{Quantization conditions and resonance parameter} of the main text are shown in~\cref{fig:rho}.

%%%%%%%%%
%%%%%%%%%
%%%%%%%%%
%%%%%%%%%

\end{document}